\documentclass{IEEEtran}
 \usepackage{cite}
\usepackage{psfrag}
\usepackage{latexsym, amsmath, subfigure,color, amsfonts, amssymb,graphicx}
\usepackage{algorithm}
\usepackage{algorithmic}
\usepackage[hidelinks]{hyperref}
\usepackage{url}
\usepackage{epstopdf}

\newtheorem{theorem}{Theorem}[section]

\newtheorem{remark}[theorem]{Remark}

\newtheorem{definition}[theorem]{Definition}

\begin{document}

\title{Privacy-preserving Average Consensus: Privacy Analysis and Optimal Algorithm Design}
\author{Jianping He$^{1, 2}$, Lin Cai$^2$, Chengcheng Zhao$^3$, Peng Cheng$^3$, and Xinping Guan$^1$
\thanks{$1$: Department of Automation, Shanghai Jiao Tong University, and Key Laboratory of System Control and Information Processing, Ministry of Education of China, Shanghai, China {\tt\small jianpinghe.zju@gmail.com}}
\thanks{$2$: The Dept. of Electrical \& Computer Engineering at the University of Victoria, BC, Canada {\tt\small{jphe@uvic.ca}; \tt\small{jianpinghe.zju@gmail.com}; \tt\small cai@ece.uvic.ca}}
\thanks{$3$: The State Key Lab. of Industrial Control Technology and Innovation Joint Research Center for Industrial Cyber Physical Systems, Zhejiang University,  China {\tt\small{zccsq90@gmail.com}; \tt\small{pcheng@iipc.zju.edu.cn}}}
}

\maketitle
\begin{abstract}
Privacy-preserving average consensus aims to guarantee the privacy of initial states and asymptotic consensus on the exact average of the initial value. In existing work, it is achieved by adding and subtracting variance decaying and zero-sum random noises to the consensus process. However, there is lack of theoretical analysis to quantify the degree of the privacy protection.  In this paper, we introduce the maximum disclosure probability that the other nodes can infer one node's initial state within a given small interval to quantify the privacy.  We develop a novel privacy definition, named $(\epsilon, \delta)$-data-privacy,  to depict the relationship between maximum disclosure probability and estimation accuracy. Then, we prove that the general privacy-preserving average consensus (GPAC) provides $(\epsilon, \delta)$-data-privacy, and provide the closed-form expression of the relationship between $\epsilon$ and $\delta$. Meanwhile, it is shown that the added noise with uniform distribution is optimal in terms of achieving the highest $(\epsilon, \delta)$-data-privacy. We also prove that when all information used in the consensus process is available, the privacy will be compromised. Finally, an optimal privacy-preserving average consensus (OPAC) algorithm is proposed to achieve the highest $(\epsilon, \delta)$-data-privacy and avoid the privacy compromission.  Simulations are conducted to verify the results.
\end{abstract}


\IEEEpeerreviewmaketitle

\section{{Introduction}} \label{Intro}
Consensus has attracted extensive attention over the past decades, since it is an efficient algorithm for distributed computing and control. A consensus algorithm refers to  the action that nodes in the network reach a global agreement regarding a certain opinion using their local neighbors' information only \cite{olfati2007consensus}\@.  Due to the strong robustness and scalability, consensus has been applied in a variety of areas, e.g., coordination and cooperation \cite{blondel2005convergence,ieeeCSM07}, distributed estimation and optimization \cite{pasqualetti2010distributed,mateos2009distributed}, sensor fusion \cite{xiao2005scheme},  distributed energy management \cite{zhao2015tsg} and sensing scheduling\cite{hetsp15},  and time synchronization \cite{schenato2011average, he2011times, carli14tac}.

Average consensus is the most commonly adopted consensus algorithm, where the agreement reached by the algorithm equals the average of all nodes' initial states. For traditional average consensus algorithms, each node will broadcast its real state to neighbor nodes
during consensus process. Hence, under traditional average consensus algorithms, the state information of each node is disclosed to its neighbor nodes. However, in some applications,  the initial states of nodes are private information, which means that nodes do not want to release their real initial states to other nodes \cite{yilin15tac}. For example, consensus
algorithm is adopted in social networks for a group of members to compute the common opinion on a subject \cite{DeGroot1974}. In this application, each member may want to keep his personal opinion on the subject secret to other members. Also, in the multi-agent rendezvous
problem \cite{lin2003}, a group of nodes want to eventually rendezvous at a certain location, while the participators may not want to release their initial locations to others. This means that when the privacy is concerned, each node's real state may not be available to the other nodes, and thus the traditional consensus algorithm becomes invalid.

Recently, researchers have investigated the privacy-preserving average consensus problem, which aims to guarantee that the privacy of initial state is preserved while average consensus can still be achieved \cite{ny14tac, Manitara13, huang12, Nozari16, yilin15tac}. The basic idea is to add random noise to the real state value during the communication to protect the privacy, and then carefully design the noise adding process such that average consensus is achieved. For example, Huang et al. \cite{huang12} designed a differentially private iterative
synchronous consensus algorithm by adding independent and exponentially decaying Laplacian noises to the consensus process. Their algorithm can guarantee differential privacy. As the algorithm may converge to a random value, the exact average consensus may not be guaranteed.  Nozari et al. \cite{Nozari16} pointed out and proved that it is impossible to achieve average consensus and differential privacy simultaneously. Hence, they design a novel linear Laplacian-based
consensus algorithm, which guarantees that an unbiased estimate of the average consensus can be achieved almost surely with differential privacy.  Manitara and Hadjicostis \cite{Manitara13} proposed a privacy preserving average consensus algorithm by adding correlated noises to the consensus process. The proposed algorithm guarantees the initial state of each node cannot be perfectly inferred by the other ``malicious" nodes.  A sufficient condition is provided under which the privacy of benign agents' initial states are preserved. More recently,  Mo and Murray in \cite{yilin15tac} well addressed the privacy-preserving average consensus problem by designing a novel PPAC algorithm, where exponentially decaying and zero-sum normal noises are added to traditional consensus process. They proved that PPAC algorithm achieves the exact average consensus in the mean-square sense, and also proved  that the algorithm achieves minimum privacy breach in the sense of disclosed space. Braca et al. in \cite{Braca16} examined the interplay between learning and privacy over multi-agent consensus networks. They provided an analytical characterization of the interplay between learning and privacy for the consensus perturbing and preserving strategy, respectively.

However, there is lack of theoretical results to quantify the degree of the privacy protection and what is the relationship between the estimation accuracy and privacy. To fill this gap, in this paper,  we provide theoretical privacy analysis for the GPAC algorithm (consider the general noise adding process) in the sense of the maximum disclosure probability that other nodes can infer one node's initial state within a given small interval (a given estimation accuracy). A privacy definition, named $(\epsilon, \delta)$-data-privacy, which is first introduced in our previous work \cite{he16tacsubmit}, is exploited to depict the maximum disclosure probability.  This privacy definition reveals the relationship between privacy and estimation accuracy. We provide theoretical results to quantify the degree of the privacy preservation and demonstrate the quantitative relationship of the estimation accuracy  and the privacy under the GPAC algorithm. Based on the analysis, it is found that the noise with uniform distribution is the optimal one in terms of achieving the highest $(\epsilon, \delta)$-data-privacy, and the exact initial state of a node can be perfectly inferred, i.e., privacy is compromised, when a node has all information used in the consensus process. Hence, to solve this problem, we design a novel OPAC algorithm to achieve average consensus as well as data-privacy. The main contributions of this paper are summarized as follows\@.
\begin{enumerate}
\item[$\bullet$] We prove that the GPAC algorithm provides $(\epsilon, \delta)$-data-privacy, and obtain a closed-form expression of the relationship between the estimation accuracy and the privacy (the relationship between $\epsilon$ and $\delta$).
\item[$\bullet$]  We prove that for the added random noises, the uniform distribution is optimal in the sense that a PPAC algorithm can achieve
         the highest privacy when the mean and variance of noises are fixed.
\item[$\bullet$]  We prove that when all the information used in the consensus process are available for the estimation, the maximum disclosure probability  will converge to one, i.e., the initial state of a node is perfectly inferred. This result reveals how the exact initial state can be inferred.
\item[$\bullet$] We design a novel OPAC algorithm to achieve average consensus while guarantees the highest $(\epsilon, \delta)$-data-privacy. It is proved that OPAC algorithm converges to the exact average consensus, and avoids the privacy to be lost even if all the information used in the consensus process is available for the estimation.
\end{enumerate}

The remainder of this paper is organized as follows\@. Section \ref{sec:pre} introduces preliminary results and problem formulation.  In Section \ref{sec:mainresult},  we provide theoretical results on the degree of privacy pretection\@. The OPAC algorithm is proposed in Section \ref{sec:opav}.   Section \ref{sec:veri} verifies the main results and conclusions are given in Section \ref{sec:conclusions}\@.

\section{Preliminaries and Problem Formulation}\label{sec:pre}

The network is abstracted as an undirected and connected graph, $G = (V, E)$, where $V$ is the set of nodes and $E$ is the set of the communication links (edges) between nodes. $(i, j)\in E$ if and only if (iff) nodes $i$ and $j$ can communicate with each other.  Let $N_i$ be the neighbor set of node $i$, where $j\in N_i$ iff $(i, j)\in E$, i.e., $N_i=\{j| j\in V, (i, j)\in E, j\neq i\}$.
\subsection{Average Consensus}
Suppose that there are $n$ ($n\geq 3$) nodes in the network (i.e., $|V|=n$), and each node $i$ has an initial scalar state $x_i(0)$, where $x_i(0)\in R$. For an average consensus algorithm, each node will communicate with its neighbor nodes and update its state based on the received information to obtain the average of all initial state's values. Hence, the traditional average consensus algorithm is given as follows,
\begin{align}\label{generalconsensus}
& x_i(k+1)=w_{ii}x_i(k)+\sum_{j\in N_i} w_{ij} x_j(k),
\end{align}
for $\forall i\in V$, which can be written in the matrix form as
\begin{align}\label{matrix_generalconsensus}
& x(k+1)=Wx(k),
\end{align}
where $w_{ii}$ and $w_{ij}$ are  weights, and $W$ is the weight matrix. It is well known from \cite{Olshevsky11} that if, 1) $w_{ii}>0$, and $w_{ij}>0$ for $(i, j)\in E$ and $w_{ij}=0$ for otherwise; and 2)  $W$ is a doubly stochastic matrix, then average consensus can be achieved by (\ref{generalconsensus}),
i.e.,
\begin{align}\label{gcconvergence}
& \lim_{k\rightarrow \infty}x_i(k)= {\sum_{\ell=1}^n x_\ell(0)\over n}=\bar{x}.
\end{align}

When the privacy of nodes' initial states are concerned,  all nodes are unwilling to release its real state to the neighbor nodes at each iteration. It means that each $x_j(k)$ is unavailable in (\ref{generalconsensus}). To preserve the privacy of  nodes' initial states,  a widely used approach is to add a random noise to the real state value when a node needs to communicate with its neighbor nodes at each iteration. We define a new state as
\begin{equation}\label{xiadd}
x_i^+(k)=x_i(k)+{\theta}_i(k), i\in V,
 \end{equation}
where ${\theta}_i(k)$ is the added random noise for privacy preservation at iteration $k$. With the noise adding process, the update equation (\ref{generalconsensus}) is changed to,
\begin{align}\label{ppaca}
&x_i(k+1)=w_{ii}x_i^+(k)+\sum_{j\in N_i} w_{ij} x_j^+(k)  \\ \label{ppacatheta}
&=w_{ii}[x_i(k)+\theta_i(k)]+\sum_{j\in N_i} w_{ij} [x_j(k)+\theta_j(k)],
 \end{align}
for $\forall i \in V$. Therefore, a privacy-preserving average consensus algorithm is to design the added noises (including the distribution and the correlations among them),  such that the goal of (\ref{generalconsensus}) is achieved under (\ref{ppaca}). Note that in (\ref{xiadd}), the noise $\theta_i(k)$ is a general random noise (where its distribution is not fixed), the algorithm (\ref{xiadd})--(\ref{ppacatheta})  is thus named as the general privacy-preserving average consensus (GPAC) algorithm in the remainder part of this paper.

\subsection{Privacy Definitions}
Under (\ref{xiadd}), the broadcast information sequence of node $i$ is $x_i^+(0), x_i^+(1), ..., x_i^+(k)$, which will be received by its neighbor nodes. Hence, each neighbor node $j$ can infer/estimate the initial state $x_i(0)$ with the received information sequence from node $i$.  Note that each of the information output, $x_i^+(k)$,  equals the weighted sum of the received information in the previous round plus a noise. Based on the information output, node $j$ will take the probability over the space of all noises $\{\theta_i(k)\}_{k=0}^\infty$ (where the space is denoted by $\Theta$) to estimate the values of the added noises. It then will infer $x_i(0)$ by using the difference between each information output and the estimated noises, i.e., $\hat{x}_i(0)= {x}_i^+({k})-\hat{\theta}_i(k)$, where $\hat{\theta}_i(k)$ is the estimation of random noise $\theta_i^k$ ($\theta_i^k={x}_i^+({k})-{x}_i(0)$). Under this estimation, we have
 \begin{equation}  \label{deprivacy}
 \Pr\left\{ |\hat{x}_i(0)-x_i(0)|\leq \epsilon\right\}=\Pr\left\{ |\hat{\theta}_i(k)-\theta_i^k|\leq \epsilon\right\},
\end{equation}
where $\epsilon\geq0$ is a small constant.

To investigate the relationship
between the estimation accuracy and privacy, by referring to  \cite{he16tacsubmit}, we then introduce an privacy definition, named $(\epsilon, \delta)$-data-privacy, where $0 \leq \epsilon$ and $0 \leq \delta\leq 1$,  as follows.
   \begin{definition}
A GPAC algorithm provides $(\epsilon, \delta)$-data-privacy, if and only if (iff),
 \begin{equation}\label{deprivacy}
\delta=\max_{\hat{\theta}_i(k)\in \Theta, k\geq 0}\Pr\{|\hat{\theta}_i(k)-\theta_i^k|\leq \epsilon\},
\end{equation}
where $\theta_i^k={x}_i^+({k})-{x}_i(0)$  and $\hat{\theta}_i(k)$ is the estimation of $\theta_i^k$.
 \end{definition}

In the above definition, the estimation
accuracy is denoted by parameter $\epsilon$ and the privacy is expressed by parameter $\delta$.  From (\ref{deprivacy}), it follows that $\delta$ is the maximum probability that each neighbor node $j$ can successfully
estimate the initial state $x_i(0)$ in a
given interval $[x_i(0)-\epsilon, x_i(0)+\epsilon]$  with the information output of node $i$ only. $\delta$ is thus named as the maximum disclosure probability.

  \begin{definition}\label{definition2}
Given an $\epsilon$, if algorithms $A_1$ and $A_2$ provide  $(\epsilon, \delta_1)$-data-privacy and $(\epsilon, \delta_2)$-data-privacy, respectively, where $\delta_1<\delta_2$, then we say that $A_1$ achieves a higher $(\epsilon, \delta)$-data-privacy than $A_2$.
 \end{definition}


\subsection{Problem Formulation}\label{sec:problem}

In this paper, we will investigate the privacy of the GPAC algorithm (\ref{xiadd})--(\ref{ppacatheta}) based on the definition of  $(\epsilon, \delta)$-data-privacy, and then design an optimal  privacy-preserving average consensus (OPAC) algorithm in terms of  $(\epsilon, \delta)$-data-privacy protection. In summary, we will consider the following four critical problems: i)  how to quantify and analyze the privacy of the GPAC algorithm;  ii) how will the distribution and correlation of the added random noises affect the privacy;  iii) when and how will a node's exact initial state be inferred by the other nodes; iv) how to achieve the optimal $(\epsilon, \delta)$-data-privacy and the exact average consensus, and avoid the privacy of nodes' initial states to be lost.




\section{Privacy Analysis of GPAC} \label{sec:mainresult}
Before presenting the main results, we first give the basic assumptions and the information set used for state estimation. Assume that the distribution and the correlation of the random variable $\theta_i(k), k=0, 1, ...$,  and the update rule of the GPAC algorithm are available to all nodes. The full topology information and $n$ are assumed to be unknown to any node, which means that each node cannot know the neighbor set of its neighbor nodes and the number of nodes in the whole network. The initial states of nodes are assumed to be independent from each other. For estimation, if there is no information of a variable, then the variable is viewed with domain $R$.  For simplicity, we assume that $\theta_i(k)$ and $\theta_j(k)$ are independently and identically distributed (i.i.d) $\forall k\geq 0$ and $i\neq j$.
Let $X$ be the output of a random variable whose distribution is unknown and with domain $R$. Without the knowledge of the distribution, according to the principle of maximum entropy, we should take the same probability over all the possible values of the random variable to estimate the values of $X$. Therefore, given an estimation $\hat{X}$, it is reasonable to assume that
\begin{equation}\label{intiassum}
\Pr\{|\hat{X}-X| \leq \epsilon \}\ll \max_{\nu\in \Theta} \int_{\nu-\epsilon}^{\nu+\epsilon} f_{\theta_i(0)} (y) \text{d} y.
\end{equation}

Then, we
define two information sets of node $i$ up to iteration $k$ as follows,
\begin{equation}
\mathcal{I}_i^0(k)=\{x_i^+(0),  ..., x_i^+(k)\},
\end{equation}
and
\begin{align}
 \mathcal{I}_i^1(k)=&\{N_i, w_{ii}, w_{ij}, x_i^+(0), x_j^+(0), \nonumber
 \\ &~~~..., x_i^+(k), x_j^+(k)|j\in N_i\}.
\end{align}
The information set $\mathcal{I}_i^0(k)$ only includes the states  $x_i^+(\ell), \ell=0, 1,..., k$, which are used for communication at iteration $\ell$. Thus, its neighbor nodes can easily obtain $\mathcal{I}_i^0(k)$ by storing the information received from node $i$ at each iteration.  The information set $\mathcal{I}_i^1(k)$ includes all  information used in consensus process (\ref{ppacatheta}) for node $i$. Other nodes may obtain these information by an eavesdropping attack.

\subsection{Privacy of the Algorithm}\label{ppacp}
In this subsection, based on the definition of $(\epsilon, \delta)$-data-privacy, we first analyze the privacy of the GPAC algorithm and reveal the relationship between the privacy and estimation accuracy, when $\mathcal{I}_i^0(k)$ is available only.

\begin{theorem}\label{corollary1}
If  $\mathcal{I}_i^0(k)$ is the only information available to node $j$ to estimate the value of $x_i(0)$ at iteration $k$, then
\begin{align} \label{deltainqnew}
\delta(k)=&\max_{\hat{\theta}_i(k)\in \Theta, k\in \mathbf{N}^+}\Pr\{|\hat{\theta}_i(k)-\theta_i^k|\leq \epsilon|_{\mathcal{I}_i^0(k)}\} \nonumber\\=& \max_{\hat{\theta}_i(0)\in \Theta}\Pr\{|\hat{\theta}_i(0)-\theta_i^0|\leq \epsilon|_{\mathcal{I}_i^0(0)}\}
\\ \label{deltainqnewad}
= & \max_{\hat{\theta}_i(0)\in \Theta} \int_{\hat{\theta}_i(0)-\epsilon}^{\hat{\theta}_i(0)+\epsilon} f_{\theta_i(0)} (y) \text{d} y,
\end{align}
i.e., the relationship between the privacy and the estimation accuracy always satisfies (\ref{deltainqnewad}), and the maximum disclosure probability is not increased with iteration.
\begin{proof}
We first prove that, under ${\mathcal{I}_i^0(0)}$, (\ref{deltainqnewad}) holds.
With ${\mathcal{I}_i^0(0)}$,  node $j$ can estimate $x_i(0)$ based on the fact that
\begin{align}\label{eq:xj0}
x_i^+(0)=x_i(0)+\theta_i(0)=x_i(0)+\theta_i^0,
\end{align}
 and the corresponding estimation $\hat{x}_i(0)$ satisfies
\begin{align}\label{eq:xje0}
\hat{x}_i(0)=x_i^+(0)-\hat{\theta}_i(0).
\end{align}
Then, for any estimation $\hat{\theta}_i(0)$, we have
\begin{align}\label{ftheta}
  & \Pr\left\{ |\hat{\theta}_i(0)-\theta_i^0|\leq \epsilon|_{\mathcal{I}_i^0(0)}\right\}\nonumber \\
=& \Pr\left\{ \theta_i(0) \in [\hat{\theta}_i(0)-\epsilon, \hat{\theta}_i(0)+\epsilon] |_{\mathcal{I}_i^0(0)}\right\}\nonumber \\
=& \int_{\hat{\theta}_i(0)-\epsilon}^{\hat{\theta}_i(0)+\epsilon} f_{\theta_i(0)|_{\mathcal{I}_i^0(0)}} (y) \text{d} y \nonumber \\ \leq& \max_{\hat{\theta}_i(0)\in \Theta} \int_{\hat{\theta}_i(0)-\epsilon}^{\hat{\theta}_i(0)+\epsilon} f_{\theta_i(0)} (y) \text{d} y,
\end{align}
which means that (\ref{deltainqnewad}) holds under information ${\mathcal{I}_i^0(0)}$ at iteration $k=0$.

Then, we prove that (\ref{deltainqnewad}) holds under ${\mathcal{I}_i^0(1)}$.  With ${\mathcal{I}_i^0(1)}$, node $j$ can estimate $x_i(0)$  by using the fact of both (\ref{eq:xj0}) and the following equation for estimation,
\begin{align} \label{eq:xj1}
{x_i^+(1)\over w_{ii}}=&{x_i(1)+\theta_i(1) \over w_{ii}}\nonumber
  \\=&x_i^+(0)+\sum_{l\in N_i} {w_{il}\over w_{ii}} x_l^+(0)+{1\over w_{ii}}\theta_i(1)\nonumber
  \\=&x_i(0)+ \theta_i(0)+{1\over w_{ii}}\theta_i(1)+\sum_{l\in N_i} {w_{il}\over w_{ii}} x_l^+(0).
\end{align}
Using (\ref{eq:xj0}) only, we have
\begin{align}\label{ftheta1d}
  & \Pr\left\{ |\hat{\theta}_i(0)-\theta_i^0|\leq \epsilon|_{\mathcal{I}_i^0(1)}\right\}\nonumber \\
 =& \int_{\hat{\theta}_i(0)-\epsilon}^{\hat{\theta}_i(0)+\epsilon} f_{\theta_i(0)|_{\mathcal{I}_i^0(1)}} (y) \text{d} y \nonumber \\
 \leq& \max_{\hat{\theta}_i(0)\in \Theta} \int_{\hat{\theta}_i(0)-\epsilon}^{\hat{\theta}_i(0)+\epsilon} f_{\theta_i(0)} (y) \text{d} y.
\end{align}
Then,  we consider the estimation  using (\ref{eq:xj1}) only.  Let
 \begin{align}\label{unstae}
 {\theta}_i^1&=\theta_i(0)+{1\over w_{ii}}\theta_i(1)+\sum_{l\in N_i} {w_{il}\over w_{ii}} x_l^+(0)\nonumber
\\& =\theta_i(0)+{1\over w_{ii}}\theta_i(1)+ \theta_i^1(1)\nonumber
\\&=\theta_i^1(0)+ \theta_i^1(1).
 \end{align}
 For any estimation $\hat{\theta}_i(1)$ (the estimation of ${\theta}_i^1$),  we have
 \begin{align}\label{ftheta1}
  & \Pr\left\{ |\hat{\theta}_i(1)-\theta_i^1|\leq \epsilon|_{\mathcal{I}_i^0(1)}\right\}\nonumber \\
\leq & \Pr\left\{| \theta_i^1-\hat{\theta}_i(1)|\leq\epsilon |_{\mathcal{I}_i^0(1), w_{ii},  \theta_i(1), \theta_i(0)}\right\}\nonumber \\
 \leq & \Pr\left\{ |\theta_i^1-\theta_i^1(0) -\hat{\theta}_i(1)+\theta_i^1(0)|\leq \epsilon |_{\mathcal{I}_i^0(1), w_{ii}, \theta_i^1(0)}\right\}\nonumber \\
 \leq & \Pr\left\{ |\theta_i^1(1) -\hat{\theta}_i^1(1)|\leq \epsilon |_{\mathcal{I}_i^0(1), w_{ii}}\right\},
\end{align}
where $\hat{\theta}_i^1(1)=\hat{\theta}_i(1)-\theta_i^1(0)$ can be viewed as one of the estimation of $\theta_i^1(1)$.
Since the initial states of nodes are independent from each other and the topology information is not available for estimating/inferring, there is at least one variable included in $\theta_i^1(1)$ which is unknown to the other nodes. Hence, $\theta_i^1(1)$ is viewed as a random variable in (\ref{ftheta1})  and its distribution is not available to the estimation. It follows that
 \begin{align}\label{ftheta11}
  & \Pr\left\{ |\hat{\theta}_i(1)-\theta_i^1|\leq \epsilon|_{\mathcal{I}_i^0(1)}\right\}\nonumber \\
\leq &\Pr\left\{ |\theta_i^1(1) -\hat{\theta}_i^1(1)|\leq \epsilon |_{\mathcal{I}_i^0(1), w_{ii}}\right\} \nonumber \\
\leq &\max_{z \in \Theta} \int_{z-\epsilon}^{z+\epsilon} f_{\theta_i(0)} (y) \text{d} y,
\end{align}
where we have used the assumption (\ref{intiassum}).
Meanwhile, note that one node can combine  (\ref{eq:xj0}) and  (\ref{eq:xj1}) together for estimation. In this case, we have
\begin{align} \label{eq:pre2}
&\Pr\{\hat{x}_i(0) \in [x_i(0)-\epsilon, x_i(0)+\epsilon]|_{\mathcal{I}_i^0(1)}\} \nonumber \\
 \leq & \max_{t_1, t_2 \in \Theta}\int_{t_1-\epsilon}^{t_1+\epsilon} \int_{t_2-\epsilon}^{t_2+\epsilon} f_{{\theta}_i^0, {\theta}_i^1} (y, z) d z \text{d} y  \nonumber \\
 \leq & \max_{t_1, t_2 \in \Theta}\int_{t_1-\epsilon}^{t_1+\epsilon} \int_{t_2-\epsilon}^{t_2+\epsilon} f_{{\theta}_i^1|{\theta}_i^0} (z|y)  f_{{\theta}_i^0} (y) d z \text{d} y
 \nonumber \\
 \leq & \max_{t\in \Theta}\int_{t-\epsilon}^{t +\epsilon}   f_{{\theta}_i(0)} (y)  \text{d} y.
\end{align}
From (\ref{ftheta1d}),  (\ref{ftheta11}) and (\ref{eq:pre2}), one concludes that (\ref{deltainqnewad}) holds under information $\mathcal{I}_i^0(1)$ at iteration $k=1$.

Following the similar analysis, we can prove that  (\ref{deltainqnewad}) holds under information set $\mathcal{I}_i^0(k)$ at any iteration $k$.  It means that  $\delta(k)$ is not an increased function of iteration $k$, although there are more information of $\mathcal{I}_i^0(k)$ than $\mathcal{I}_i^0(0)$ for $k>0$.

We thus have completed the proof.
\end{proof}
\end{theorem}

From the above proof, it is observed that the privacy is not decreased with iteration when only the information set $\mathcal{I}_i^0$ ($=\{\mathcal{I}_i^0(k)| k=0, 1, ..., \infty\}$) is available for estimation. The main reason is that based on $\mathcal{I}_i^0$, node $j$ cannot know the neighbor set information of node $i$, so that after one iteration there are unknown information embedded into  $x_i^+(k)$ for $k\geq 1$.  Hence, after one iteration, using $x_i^+(k)$ for $k\geq 1$ cannot improve the estimation accuracy.  Also, one can see that the value of $\delta$ does not depend on the estimation approaches.  Hence,  we state the following theorem.
{\begin{theorem}\label{theorem:4.1}
If $\mathcal{I}_i^0$ is the only information available to the other nodes to estimate the value of $x_i(0)$, the GPAC algorithm achieves $(\epsilon, \delta)$-data-privacy, where $\epsilon$ and $\delta$ satisfy
\begin{equation}\label{deltainq}
\delta= \max_{\hat{\theta}_i(0)\in \Theta} \int_{\hat{\theta}_i(0)-\epsilon}^{\hat{\theta}_i(0)+\epsilon} f_{\theta_i(0)} (y) \text{d} y
\end{equation}
and $\lim_{\epsilon\rightarrow 0}\delta=0$.
\end{theorem}}

\begin{remark}
It should be noticed that the results in the above two theorems are obtained under the assumption that the topology information is unknown to the nodes. If the assumption is relaxed, the above results could not be true for the GPAC algorithm in some cases. For example, if the topology information is available and $N_i\subseteq N_j$, then $x_i^+(0)$ and $\sum_{l\in N_i} {w_{il}\over w_{ii}} x_l^+(0)={w_{ij}\over w_{ii}} x_j^+(0)$ in (\ref{eq:xj1}) are available to node $j$.  It leads to that the value of ${\theta}_i(1)$ is released, which may decrease the uncertainty of $\theta_i(0)$ due to the correlation between them. Then, $f_{\theta_i(0)|_{\mathcal{I}_i^0(1)}} (y)$  in (\ref{ftheta1d}) will have a smaller variance than $f_{\theta_i(0)}(y)$, such that $\delta$  increases w.r.t. $k$ in this case. Therefore,  (\ref{deltainqnew}) and  (\ref{deltainqnewad}) are no longer guaranteed. 
\end{remark}

From the above theorem, one sees that $\delta$ depends only on $f_{\theta_i(0)} (y)$ and $\epsilon$ since the estimation $\hat{\theta}_i(0)$ can be any value in the domain of $\theta_i(0)$. Thus, $\delta$ is a function of  $f_{\theta_i(0)} (y)$ and $\epsilon$, i.e, $\delta=\delta(f_{\theta_i(0)} (y), \epsilon)$.  Based on Definition \ref{definition2}, a smaller $\delta$ can provide a higher $(\epsilon, \delta)$-data-privacy for any given $\epsilon$. Then, we aim to find the optimal distribution for $\theta_i(0)$ such that the algorithm can achieve the highest $(\epsilon, \delta)$-data-privacy.

\subsection{Optimal Noise Distribution}
In this subsection, we find an optimal distribution for the noise adding process in the sense of achieving the highest $(\epsilon, \delta)$-data-privacy for the GPAC algorithm. Note that a smaller $\epsilon$ means a higher accuracy estimation. It means that when $\epsilon$ becomes smaller, the value of $\delta$ is more important for the privacy preservation. Hence, we define the optimal distribution for privacy concerns as follows.
 \begin{definition} \label{definition3}
 Let $f_{\theta_i(0)}^*(y)$ be the optimal distribution of $\theta_i(0)$, it means that for any given distribution $f_{\theta_i(0)}^1 (y)$, there exists an $\epsilon_1$ such that $\delta(f_{\theta_i(0)}^* (y), \epsilon)<\delta(f_{\theta_i(0)}^1 (y), \epsilon)$ holds for $\forall \epsilon\in (0, \epsilon_1]$.
 \end{definition}

To obtain the optimal distribution described in Definition \ref{definition3}, we define $\arg\min_{f_{\theta_i(0)} (y)}  \delta=f_{\theta_i(0)}^*(y)$. Then, we formulate the following minimization problem,
\begin{equation}
 \begin{split}\label{problem:p1}
\min_{f_{\theta_i(0)} (y)} & ~~ \delta
\\ s.t. 
 ~~ &\mathbf{E} \{\theta_i(0)\}=0,\\
 ~~ & \mathbf{Var} \{\theta_i(0)\}=\sigma^2.
\end{split}
\end{equation}
The solution of (\ref{problem:p1}) is the optimal distribution for the added noises with a given mean and variance in terms of $(\epsilon, \delta)$-data-privacy for the GPAC algorithm. 
\begin{theorem}
If  $\mathcal{I}_i^0$ is the only information available to node $j$ to estimate the value of $x_i(0)$, then the optimal solution of  problem (\ref{problem:p1}) is that
\begin{equation} \label{oftheta}
f_{\theta_i(0)}^* (y)=\left\{ \begin{aligned}
        &{ 1 \over 2\sqrt{3}\sigma},  && \textsf{if}~ y\in [-  \sqrt{3}\sigma, \sqrt{3}\sigma], \\
       &0,  && \textsf{otherwise},
                          \end{aligned} \right.
                          \end{equation}
i.e., given the finite variance, the uniform distribution is optimal.

\begin{proof}
We prove this theorem by contradiction. Without loss of generality, we assume that $\sigma^2={1\over 3}$.  Let $f_1(y)$ and $f_2(y)$ be the PDF of two random variables with mean $0$ and variance $\sigma^2={1\over 3}$, and they follow a uniform and non-uniform distribution, respectively. Clearly, we have
\begin{equation} \label{f1}
f_1 (y)=\left\{ \begin{aligned}
        &{1 \over 2},  && \textsf{if}~ y\in [- 1, 1], \\
       &0,  && \textsf{otherwise}.
                          \end{aligned} \right.
                          \end{equation}

Suppose that the non-uniform distribution $f_2(y)$ is the optimal distribution. From Definition \ref{definition3},  there exists an $\epsilon_2$,  such that
\begin{align} \label{deltaequniform}
\max_{t\in R} \int_{t-\epsilon}^{t+\epsilon} f_1 (y) \text{d} y>\max_{t\in R} \int_{t-\epsilon}^{t+\epsilon} f_2 (y) \text{d} y,
\end{align}
holds for $\forall \epsilon \in (0, \epsilon_2]$. Since the above equation holds for arbitrarily small value of $\epsilon$, we infer that
\[\max_{y\in R}  f_1 (y)>\max_{y\in R}  f_2 (y).\]
Since $f_1 (y)$ is a uniform distribution satisfying (\ref{f1}),
\[ f_1 (y)- f_2 (y)>0, ~ y\in [-1, 1].\]
It directly follows that
\begin{align} \label{probadefinition}
\int_{-1}^{1} f_1 (y)\text{d} y-\int_{-1}^{1} f_2(y) \text{d} y > 0.
\end{align}
From the definition of a PDF,  we have $\int_{-1}^{1} f_1 (y)\text{d} y=1$. Then, we infer from (\ref{probadefinition}) that
\begin{align} \label{main1}
\int_{-1}^{1} f_2(y) \text{d} y<1.
\end{align}

Since both  $f_1(y)$ and $f_2(y)$ have mean $0$ and variance $\sigma^2={1\over 3}$, we have
\begin{align} \label{main22}
&\int_{-\infty}^{+\infty} f_1(y) y^2\text{d} y-\int_{-\infty}^{\infty} f_2(y) y^2\text{d} y=0,
\end{align}
which means that
\begin{align} \label{main2}
\int_{-1}^{1}\left( f_1(y) - f_2(y)\right) y^2 \text{d} y=\left(\int_{-\infty}^{-1}+\int_{1}^{+\infty}\right)f_2(y) y^2\text{d} y.
\end{align}
For the left hand side of (\ref{main2}),  we have
\begin{align} \label{main3}
  \int_{-1}^{1}\left( f_1(y) - f_2(y)\right) y^2 \text{d} y&<\int_{-1}^{1}\left( f_1(y) - f_2(y)\right)\text{d} y \nonumber
  \\&=1- \int_{-1}^{1} f_2(y) \text{d} y.
\end{align}
For the right hand side of (\ref{main2}), since we have $\int_{-\infty}^{+\infty}  f_2(y)\text{d} y=1$ and (\ref{main1}),  it holds that
\begin{align} \label{main4}
\left(\int_{-\infty}^{-1}+\int_{1}^{+\infty}\right)f_2(y) y^2\text{d} y &> \left(\int_{-\infty}^{-1}+\int_{1}^{+\infty}\right)f_2(y) \text{d} y \nonumber
  \\&=1- \int_{-1}^{1} f_2(y) \text{d} y.
\end{align}

Combining (\ref{main2}),  (\ref{main3}) and  (\ref{main4}) renders a contradiction that
\begin{align}
1- \int_{-1}^{1} f_2(y) \text{d} y&<\int_{-1}^{1}\left( f_1(y) - f_2(y)\right) y^2 \text{d} y\nonumber \\&<1- \int_{-1}^{1} f_2(y) \text{d} y.
\end{align}
Hence, we cannot find a non-uniform distribution $f_2(y)$ such that the value of $\delta$ is smaller than that under uniform distribution $f_1(y)$. It means that, given the finite variance, the uniform distribution is the optimal solution of (\ref{problem:p1}). Then, based on the definition of uniform distribution, it is not difficult to obtain (\ref{oftheta}).

We thus have completed the proof.
\end{proof}

\end{theorem}

We have known that for the existing PPAC algorithm proposed in \cite{yilin15tac}, the normal distribution noises is used in the noise adding process. It follows from Theorem \ref{theorem:4.1} that the PPAC algorithm provides $(\epsilon, \delta)$-data-privacy with
 \begin{equation*}
\delta= {1 \over \sigma \sqrt{2 \pi}} \int_{-\epsilon}^{\epsilon} \exp\left(-{y^2\over 2 \sigma^2}\right)\text{d} y.
\end{equation*}
If we use the uniform distribution noises to substitute the normal distribution noises, it can still provide $(\epsilon, \delta)$-data-privacy, where $\delta={\epsilon\over  \sqrt{3}\sigma}$.
Clearly, given a small $\epsilon$ ($\ll  \sigma^2$), we have
\[{\epsilon\over  \sqrt{3}\sigma}< {1 \over \sigma \sqrt{2 \pi}} \int_{-\epsilon}^{\epsilon} \exp\left(-{y^2\over 2 \sigma^2}\right)\text{d} y,\]
which means that the privacy of PPAC is enhanced.

\subsection{Privacy Compromission}
In this subsection, we reveal that for the GPAC algorithm, when $\mathcal{I}_i^1(k)$ (including more information than $\mathcal{I}_i^0(k)$, e.g., the topology information and information used in consensus process) is available to other nodes for estimation,  the exact initial state of node $i$ can be perfectly inferred, and thus the privacy of the initial state is compromised.

\begin{theorem} \label{theorem4.5}
If the information set $\mathcal{I}_i^1(k)$ of node $i$ is available to the other nodes for estimation, then
\begin{align}\label{cprob1}
\delta(k)\geq  \max_{\hat{\theta}_i(0)\in \Theta} \int_{\hat{\theta}_i(0)-\epsilon}^{\hat{\theta}_i(0)+\epsilon} f_{\theta_i(0)| \theta_i(1), ..., \theta_i(k)} (y) \text{d} y, \forall k\geq 0,
\end{align}
where $f_{\theta_i(0)| \theta_i(1), ..., \theta_i(k)} (y)$ is the conditional PDF of $\theta_i(0)$ given conditions $\theta_i(1), ..., \theta_i(k)$.  Then, if $\sum_{\ell=0}^\infty \theta_i(\ell)=0$, we have $\delta=1$, i.e., $x_i(0)$ is disclosed and the privacy is compromised.

\begin{proof}
Based on the information set $\mathcal{I}_i^1(k)$, the information of weights and states used in (\ref{ppaca}) is available. That is the state sequence  $x_i(1), x_i(2), ..., x_i(k)$  of node $i$ is released to   other nodes. Then, with (\ref{xiadd}), one obtains the values of  $\theta_i(1), \theta_i(2), ...., \theta_i(k)$. Thus, when $k>0$, all the adding noises and the states of node $i$ are available to other nodes, except $x_i(0)$ and $\theta_i(0)$.

Then, under information set $\mathcal{I}_i^1(k)$, using (\ref{eq:xj0}), we have
\begin{align}\label{ftheta1add}
  & \Pr\left\{ |\hat{\theta}_i(0)-\theta_i^0|\leq \epsilon|_{\mathcal{I}_i^0(k)}\right\}\nonumber \\
 =& \int_{\hat{\theta}_i(0)-\epsilon}^{\hat{\theta}_i(0)+\epsilon} f_{\theta_i(0)|_{\mathcal{I}_i^0(k)}} (y) \text{d} y \nonumber \\
 =& \int_{\hat{\theta}_i(0)-\epsilon}^{\hat{\theta}_i(0)+\epsilon} f_{\theta_i(0)|\theta_i(1), ..., \theta_i(k)} (y) \text{d} y.
\end{align}
According the definition of $\delta$, it follows that
\begin{align}\label{dtfc}
 \delta(k)\geq & \max_{\hat{\theta}_i(0)\in \Theta} \Pr\left\{ |\hat{\theta}_i(0)-\theta_i^0|\leq \epsilon|_{\mathcal{I}_i^0(k)}\right\}\nonumber \\
 \geq &  \max_{\hat{\theta}_i(0)\in \Theta} \int_{\hat{\theta}_i(0)-\epsilon}^{\hat{\theta}_i(0)+\epsilon} f_{\theta_i(0)|\theta_i(1), ..., \theta_i(k)} (y) \text{d} y,
\end{align}
which means that (\ref{cprob1}) holds.

When $\sum_{\ell=0}^\infty \theta_i(\ell)=0$, we have
\begin{align}\label{thetaavai}
\theta_i(0)=-\sum_{\ell=1}^\infty \theta_i(\ell).
\end{align}
Since  $\theta_i(1), \theta_i(2), ...., \theta_i(k)$ are available under $\mathcal{I}_i^1(k)$ for any integer $k$, $\theta_i(0)$ is inferred with (\ref{thetaavai}) when $k \rightarrow \infty$, i.e., $\theta_i(0)$ is fixed and no longer a random variable given $\theta_i(1), \theta_i(2), ...., \theta_i(\infty)$. It follows that \[\lim_{k \rightarrow \infty}\max_{\hat{\theta}_i(0)\in \Theta} \int_{\hat{\theta}_i(0)-\epsilon}^{\hat{\theta}_i(0)+\epsilon} f_{\theta_i(0)|\theta_i(1), ..., \theta_i(k)} (y) \text{d} y=1,\]
which implies that $\delta=1$.
Actually, when both $x_i^+(0)$ and $\theta_i(0)$ in  (\ref{eq:xj0}) are disclosed, $x_i(0)$ is disclosed.

We thus have completed the proof.
\end{proof}

\end{theorem}

Consider the existing PPAC algorithms, e.g., \cite{yilin15tac, he16tacsubmit}.
One obtain the correlation of the added noises satisfies
\begin{align}\label{tck}
\sum_{\ell=0}^k \theta_i(\ell) &= \theta_i(0)+ \sum_{\ell=1}^k
\left[\varrho^\ell \nu_i(\ell)-\varrho^{\ell-1} \nu_i(\ell-1)\right]
\nonumber \\&=\nu_i(0)-\varrho^{0} \nu_i(0)+\varrho^1 \nu_i(1)-\varrho^{1} \nu_i(1)+\varrho^2 \nu_i(2) \nonumber \\&~~~- ... -\varrho^{k-1} \nu_i(k-1)+\varrho^k \nu_i(k)
\nonumber \\&= \varrho^k \nu_i(k) = \phi_i(k),
\end{align}
where $\nu_i(k)$ is a random variable with fixed mean ($=0$) and variance ($=\sigma^2$). Given $\theta_i(1), ..., \theta_i(k)$, we obtains that
$\theta_i(0)= \phi_i(k)-\sum_{\ell=1}^k \theta_i(\ell)$, where $\sum_{\ell=1}^k \theta_i(\ell)$ is known.  Then, we have
\begin{align}\label{pri1k}
 & \Pr\left\{|\hat{\theta}_i(0)-\theta_i(0)|\leq \epsilon |_{\mathcal{I}_i^1(k)} \right\} \nonumber \\
= & \Pr\left\{ |\hat{\phi}_i(k)-\phi_i(k)|\leq \epsilon \right\}\nonumber \\
=& \int_{\hat{\phi}_i(k)-\epsilon}^{\hat{\phi}_i(k)+\epsilon} f_{\phi_i(k)} (y) \text{d} y,
\end{align}
and
 \begin{align*}
 &\max_{\hat{\theta}_i(0)\in \Theta} \int_{\hat{\theta}_i(0)-\epsilon}^{\hat{\theta}_i(0)+\epsilon} f_{\theta_i(0)| \theta_i(1), ..., \theta_i(k)} (y) \text{d} y\nonumber\\&= \max_{\hat{\theta}_i(0)\in \Theta} \int_{\hat{\phi}_i(k)-\epsilon}^{\hat{\phi}_i(k)+\epsilon} f_{\phi_i(k)} (y) \text{d} y,
 \end{align*}
which satisfies (\ref{cprob1}). When $k\rightarrow \infty$, we have
 \[\lim_{k\rightarrow \infty}\max_{\hat{\theta}_i(0)\in \Theta} \int_{\hat{\phi}_i(k)-\epsilon}^{\hat{\phi}_i(k)+\epsilon} f_{\phi_i(k)} (y) \text{d} y=1,\]
 since the variance of $\hat{\phi}_i$ satisfies $\lim_{k \rightarrow \infty}\varrho^{2k} \sigma^2=0$, and thus $\delta=1$. Therefore, it further verifies the result given in Theorem \ref{theorem4.5}.

\subsection{Further Discussion on Privacy}
Differential privacy is a well-known and widely used privacy concept in computer and communication area \cite{Dwork06}, and it has been employed in control and network systems recently \cite{Cortes16}. A differentially private algorithm promises that any two similar/close inputs will have approximately the same outputs, so that 
an adversary cannot infer from the data output with a high probability whether the data  are associated with a single user or not. 
It has been proved by Nozari et al. in \cite{Nozari16} that nodes in the network system cannot simultaneously converge to the average of their initial states and preserve differential privacy of their initial states. This motivated us to develop the definition of the $(\epsilon, \delta)$-data-privacy. The proposed $(\epsilon, \delta)$-data privacy can be used to reveal the
relationship between the the maximum data disclosure
probability ($\delta$) under a given estimation accuracy
range ($\epsilon$).

Consider the general noise adding mechanism that added a random noise to the initial data for data publishing. It is well known that when the adding noise is Laplacian noise, the mechanism ensures differential privacy, but if the noise is Gaussian or Uniform distribution, the differential privacy cannot be guaranteed. Hence, the uniform noise is not good in the sense of differential privacy. However,  in term of $(\epsilon, \delta)$-data-privacy, it is shown in this paper that both the Gaussian and Uniform noise are $(\epsilon, \delta)$-data-private, and using the Uniform noise can achieve the highest privacy. Clearly, the privacy of $(\epsilon, \delta)$-data-privacy is different from that of differential privacy. It is worth to investigate the relationship between these two kinds of privacy definition in theory, which beckons further investigation.

\section{OPAC Algorithm}\label{sec:opav}
In this section, we design an OPAC algorithm to achieve the highest $(\epsilon, \delta)$-data-privacy, and at the same time to avoid privacy to be compromised even if the information $\mathcal{I}_i^1(\infty)$ of each node $i$ is available to other nodes.

\subsection{Algorithm Design}
From the privacy analysis in the above section, we note that the uniform distribution is optimal for the added noise in terms of achieving the highest $(\epsilon, \delta)$-data-privacy with $\delta= {\epsilon\over  \sqrt{3}\sigma}$ (given variance $\sigma$). Hence, in each iteration of the OPAC algorithm, we will use uniform distribution noise. We also note that privacy is compromised when $\mathcal{I}_i^1(\infty)$ is available. It is because that the nodes can use $\mathcal{I}_i^1(\infty)$ to obtain the real values of $\theta_i(1), \theta_i(2), ...., \theta_i(\infty)$, and then use the correlation $\sum_{k=0}^\infty\theta_i(k)=0$ to infer $\theta_i(0)$, and thus the value of $x_i(0)$ is revealed. To avoid the privacy compromission in this case, we introduce a secret continuous function $F_{ij}(z): R\rightarrow R$ for node $i$ with respect to its neighbor node $j$. Suppose that $F_{ij}(z)$ and $F_{ji}(z)$ are only available to nodes $i$ and $j$, and $F_{ij}(z)$ may or may not equal to $F_{ji}(z)$. Then, the OPAC algorithm is described as follows.

\begin{algorithm}
\caption{: \emph{OPAC Algorithm}} \label{SCDA}
{\small{
\begin{algorithmic}[1]
\STATE \textbf{Initialization:} Each node $i$ selects a uniform distribution random variable $\nu_i(0)$ from interval $[- \sqrt{3}\sigma, \sqrt{3}\sigma]$, and arbitrarily selects a constant sequence $z_{ij}$ ($\in R$) for $j\in N_i$.
\STATE Let $\mathbf{\theta}_i(0)=\nu_i(0)$ and $\mathbf{x}_i^+(0) = \mathbf{x}_i(0) +\theta_i(0)$. Then, each node $i$ transmits $\mathbf{x}_i^+(0)$ and $z_{ij}$ to its neighbor node $j$.
\STATE  Each node $i$ calculates $\tilde{\nu}_i(0)$ by
\begin{align}\label{tnu}
\tilde{\nu}_i(0)=\nu_i(0)-\sum_{j\in N_i}\left[F_{ij}(z_{ij})-F_{ji}(z_{ji})\right], \forall i \in V.
\end{align}
\STATE \textbf{Iteration:}  Each node updates its state with (\ref{ppaca}).
\STATE Each node generates a uniform distribution random variable $\nu_i(k)$ from interval $ [- \sqrt{3}\sigma, \sqrt{3}\sigma]$ for $k\geq 1$.
\STATE Each node $i$ uses $\theta_i(k)$ in (\ref{xiadd}) to get $x_i^+(k)$, where
\begin{equation} \label{thetaad}
\theta_i(k)=\left\{ \begin{aligned}
        & \varrho \nu_i(1)- \tilde{\nu}_i(0), && \textsf{if}~ k=1; \\
       &\varrho^k \nu_i(k)-\varrho^{k-1} \nu_i(k-1),  &&  \textsf{if}~ k\geq 2,
                          \end{aligned} \right.
                          \end{equation}
where $\varrho\in (0, 1)$ is a constant for all nodes.
\STATE Each node $i$ communicates with its neighbors with $x_i^+(k)$.
\STATE  Let $k=k + 1$ and go to step 4.
\end{algorithmic} } }
\end{algorithm}

\subsection{Convergence and Privacy Analysis}
In this subsection, we analyze the convergence and the privacy of the OPAC algorithm.
\begin{theorem}\label{theorem:c}
Using the OPAC algorithm, we have (\ref{gcconvergence}) holds for $\forall i\in V$, i.e., an exact average consensus is achieved.

\begin{proof}
From Theorem 4.1 of \cite{he16tacsubmit},  we know that if the added noises in (\ref{xiadd}) are bounded and decaying, and the sum of all nodes' added noises equals zero, then average consensus can be achieved. In the following, we prove that the added noises used for the OPAC algorithm
satisfy these conditions.

We first prove that the added noises are bounded and exponentially decaying. Clearly,  $\mathbf{\theta}_i(0)=\nu_i(0)\in [- \sqrt{3}\sigma, \sqrt{3}\sigma]$ is bounded. Since each $F_{ij}(z)$ is continuous function, its value is bounded for any given $z$. Then, it follows from (\ref{tnu}) that $\tilde{\nu}_i(0)$ is bounded. For $k\geq 1$, because $\nu_i(k)$ is selected from interval $ [- \sqrt{3}\sigma, \sqrt{3}\sigma]$ and $\theta_i(k)$ is generated by (\ref{thetaad}),  it is not difficult to infer that each $\theta_i(k)$ is bounded. Meanwhile, it follows from (\ref{thetaad}) that
\begin{align*}
\lim_{k\rightarrow \infty} |\theta_i(k)|&\leq \lim_{k\rightarrow \infty} |\varrho^k \nu_i(k)-\varrho^{k-1} \nu_i(k-1) |\\
&\leq \lim_{k\rightarrow \infty}[ \varrho^k { \sqrt{3}\sigma}+\varrho^{k-1} {\sqrt{3}\sigma}]=0,
\end{align*}
which means that the noises are decaying and converge to $0$.

Next, we prove that the sum of all nodes' added noises equals to zero. Note that
\begin{align*}
\sum_{i=1}^n\sum_{k=0}^\infty \theta_i(k)&=\sum_{i=1}^n \theta_i(0)+\sum_{i=1}^n\theta_i(1)\\&+ \sum_{i=1}^n\sum_{k=2}^\infty (\varrho^k \nu_i(k)-\varrho^{k-1} \nu_i(k-1)) \\&= \sum_{i=1}^n \nu_i(0)+\sum_{i=1}^n(\varrho \nu_i(1)- \tilde{\nu}_i(0))\\&+ \sum_{i=1}^n (\varrho^\infty \nu_i(\infty)-\varrho^{1} \nu_i(1))\\&=\sum_{i=1}^n \nu_i(0)-\sum_{i=1}^n \tilde{\nu}_i(0),
\end{align*}
where we have used the fact that $\varrho^\infty \nu_i(\infty)=0$. Substituting (\ref{tnu}) into the above equation yields that
\begin{align*}
\sum_{i=1}^n\sum_{k=0}^\infty \theta_i(k)&=\sum_{i=1}^n \nu_i(0) \\&-\sum_{i=1}^n\left[\nu_i(0)-\sum_{j\in N_i}\left(F_{ij}(z_{ij})-F_{ji}(z_{ji})\right)\right]\\&=\sum_{i=1}^n \sum_{j\in N_i}\left[F_{ji}(z_{ji})-F_{ij}(z_{ij})\right].
\end{align*}
Since for each pair of $F_{ji}(z_{ji})-F_{ij}(z_{ij})$ using in node $i$, there exists a pair of $F_{ij}(z_{ij})-F_{ji}(z_{ji})$ with negative value using in node $j$, it follows that
\[\sum_{i=1}^n \sum_{j\in N_i}\left[F_{ji}(z_{ji})-F_{ij}(z_{ij})\right]=0.\]
Hence, we have $\sum_{i=1}^n\sum_{k=0}^\infty \theta_i(k)=0$.

Thus, the proof is completed.
\end{proof}

\end{theorem}

The following theorem can be obtained from Theorem \ref{theorem:4.1} directly, since OPAC is one of the GPAC algorithm.
\begin{theorem}
If  $\mathcal{I}_i^0$ is the only information available to the other nodes to estimate the value of $x_i(0)$, then the OPAC algorithm achieves $(\epsilon, \delta)$-data-privacy, where $\delta= {\epsilon\over  \sqrt{3}\sigma}$
and $\lim_{\epsilon\rightarrow 0}\delta=0$.
\end{theorem}

Then, the following theorem shows that under $\mathcal{I}_i^1$, the privacy compromission can be avoided by OPAC.
\begin{theorem}
Suppose that the information set $\mathcal{I}_i^1$ of node $i$ is available to the other nodes and each node has at least two neighbors (i.e., $|N_i|\geq 2$ for all $i\in V$). Then, the privacy compromission can be avoided by OPAC.

\begin{proof}
It has been known that when $\mathcal{I}_i^1$ of node $i$ is available to other nodes,  its neighbor node $j$ can obtain the real values of $\theta_i(1), \theta_i(2), ...., \theta_i(\infty)$. Then, the value of $\sum_{k=1}^\infty \theta_i(k)$ is released. Note that
\begin{align*}
\sum_{k=1}^\infty \theta_i(k)&=(\varrho^1 \nu_i(1)-\tilde{\nu}_i(0))+\sum_{k=2}^\infty \theta_i(k) \\&=(\varrho^1 \nu_i(1)-\tilde{\nu}_i(0))+(\varrho^\infty \nu_i(\infty)-\varrho^{1} \nu_i(1))\\&= \tilde{\nu}_i(0).
\end{align*}
It means that the value of $\tilde{\nu}_i(0)$ is released and available to node $j$. From (\ref{tnu}), one sees that $\tilde{\nu}_i(0)\neq \theta_i(0)${\footnote{This is the main difference between OPAC and PPAC algorithm, and the main reason why OPAC can avoid privacy compromission.}} and
\begin{align}\label{thetanu}
\tilde{\nu}_i(0)=\theta_i(0)-\sum_{j\in N_i}\left[F_{ij}(z_{ij})-F_{ji}(z_{ji})\right].
\end{align}
Since $|N_i|\geq 2$ and only $F_{ij}$ and $F_{ji}$ is known to node $j$, there exists $F_{ij_o}(z_{ij_o})-F_{j_oi}(z_{j_oi})$ for $j_0\in N_i$ in (\ref{thetanu}) is not known by node $j$. Meanwhile, $F_{ij_o}(z_{ij_o})-F_{j_oi}(z_{j_oi})$ has domain $R$, thus one infers that for any $c\in[- \sqrt{3}\sigma, \sqrt{3}\sigma]$,
 \[\Pr\{\theta_i(0)=c| \tilde{\nu}_i(0)\}=\Pr\{\theta_i(0)=c\}.\]
Hence, even if the value of $\tilde{\nu}_i(0)$ is released, node $j$ cannot increase the estimation accuracy of $\theta_i(0)$ with  (\ref{thetanu}).
One thus concludes that based on the OPAC algorithm, the privacy compromission is avoided.

We thus have completed the proof.
\end{proof}
\end{theorem}

If node $i$ has only one neighbor node $j$, node $j$ can infer the value of $F_{ij}(z_{ij})-F_{ji}(z_{ji})$. Then, from (\ref{thetanu}),  node $j$ can obtain the value of $\theta_i(0)$ and $x_i(0)$ when $\tilde{\nu}_i(0)$ is known.

\begin{remark}
From the above two theorems, one sees that using OPAC algorithm, we have $\delta={\epsilon\over  \sqrt{3}\sigma}$, which is the optimal privacy that can be achieved from solving problem (\ref{problem:p1}). Furthermore, $\delta={\epsilon\over  \sqrt{3}\sigma}$ can be guaranteed by OPAC algorithm under $\mathcal{I}_i^1(\infty)$.  Thus, OPAC algorithm can achieve much higher $(\epsilon, \delta)$-data-privacy than the existing PPAC.
\end{remark}

\section{Performance Evaluation}\label{sec:veri}

In this section, we conduct simulations to verify the obtained theoretical results and evaluate the performance of the proposed OPAC algorithm.

\subsection{Simulation Scenario}
Consider the network with $50$ nodes which are randomly deployed in a $100\textrm{m}\times 100\textrm{m}$ area, and the
maximum communication range of each node is $30\textrm{m}$. We consider the normal distribution and uniform distribution of the added noises, respectively, where the mean and variance of them are set $0$ and $\sigma^2=1$. We set $\varrho=0.9$. The initial states of the nodes are randomly selected from $[0, 10]$.  The function, $d(t)=\max\limits_{i \in\mathcal {V}}|x_i(t)-\bar{x}|$, is defined as the maximum difference between the nodes' states and the average value.
\subsection{Verification}
\begin{figure*}[t]
\begin{center}
\vspace*{-0pt}
\subfigure[convergence]{\label{lowpr}
\includegraphics[width=0.32\textwidth]{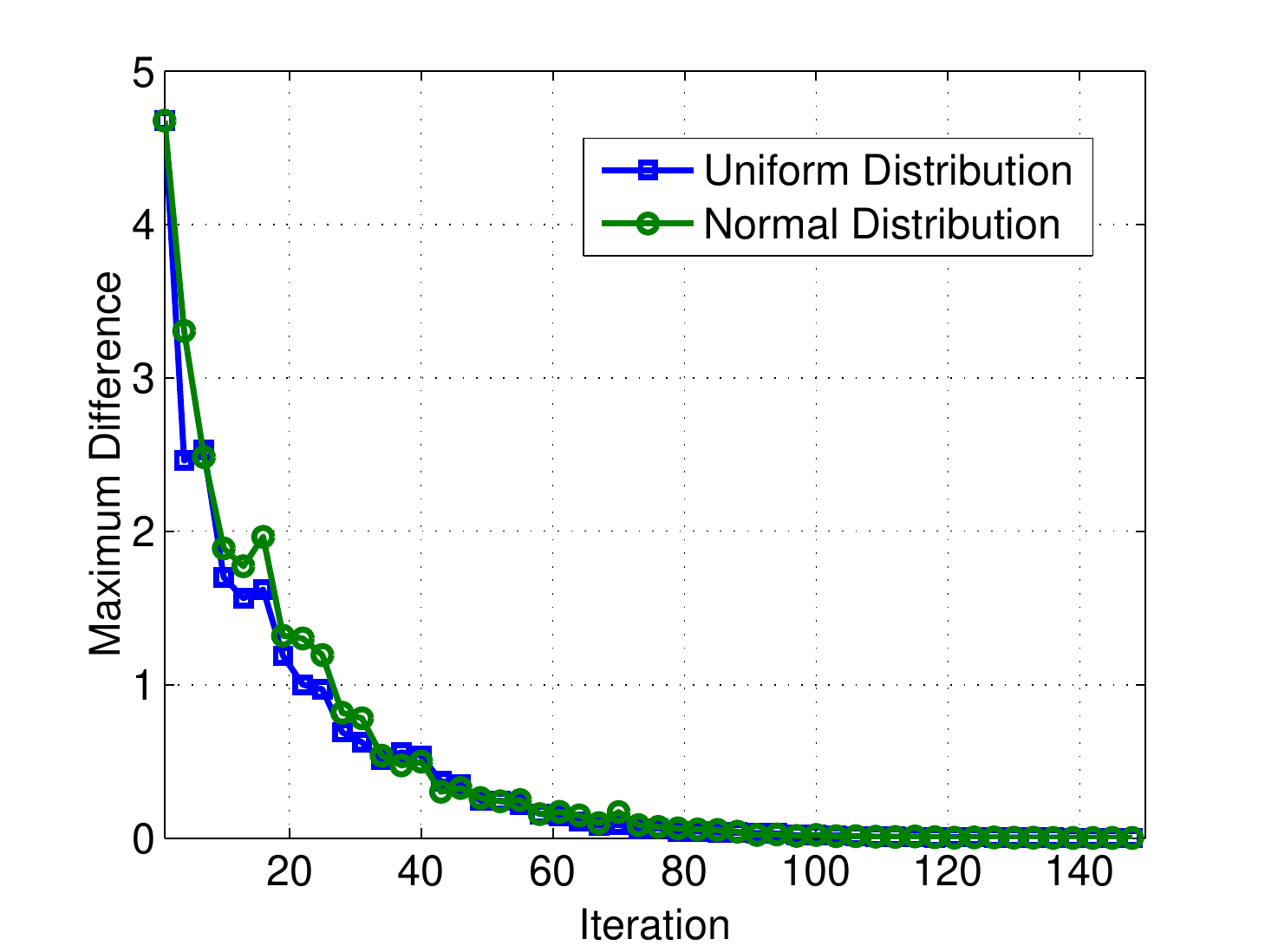}} \subfigure[privacy under $\mathcal{I}_i^0$]{\label{payment}
\includegraphics[width=0.32\textwidth]{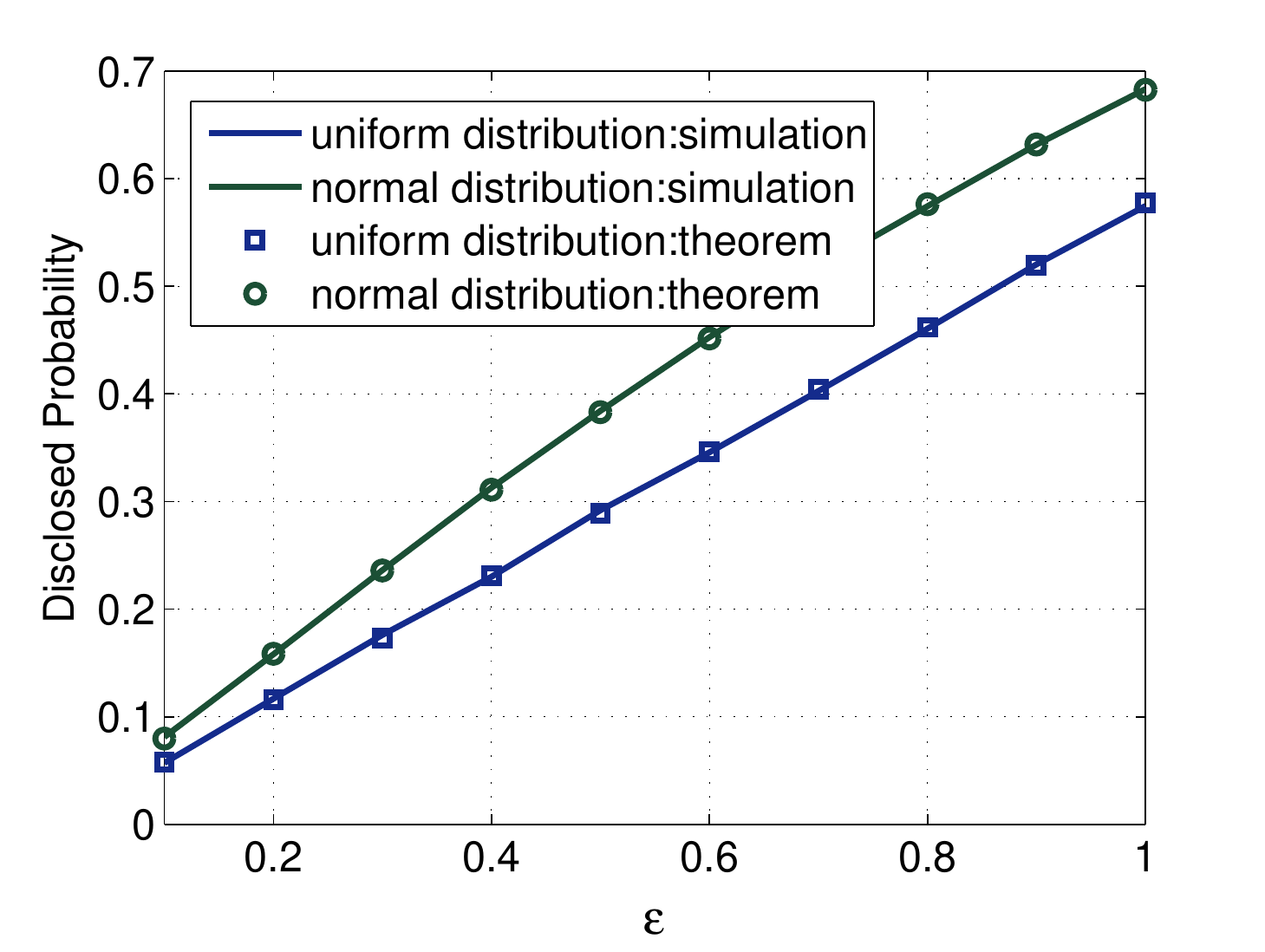}}
\subfigure[privacy under $\mathcal{I}_i^1$]{\label{demands}
\includegraphics[width=0.32\textwidth]{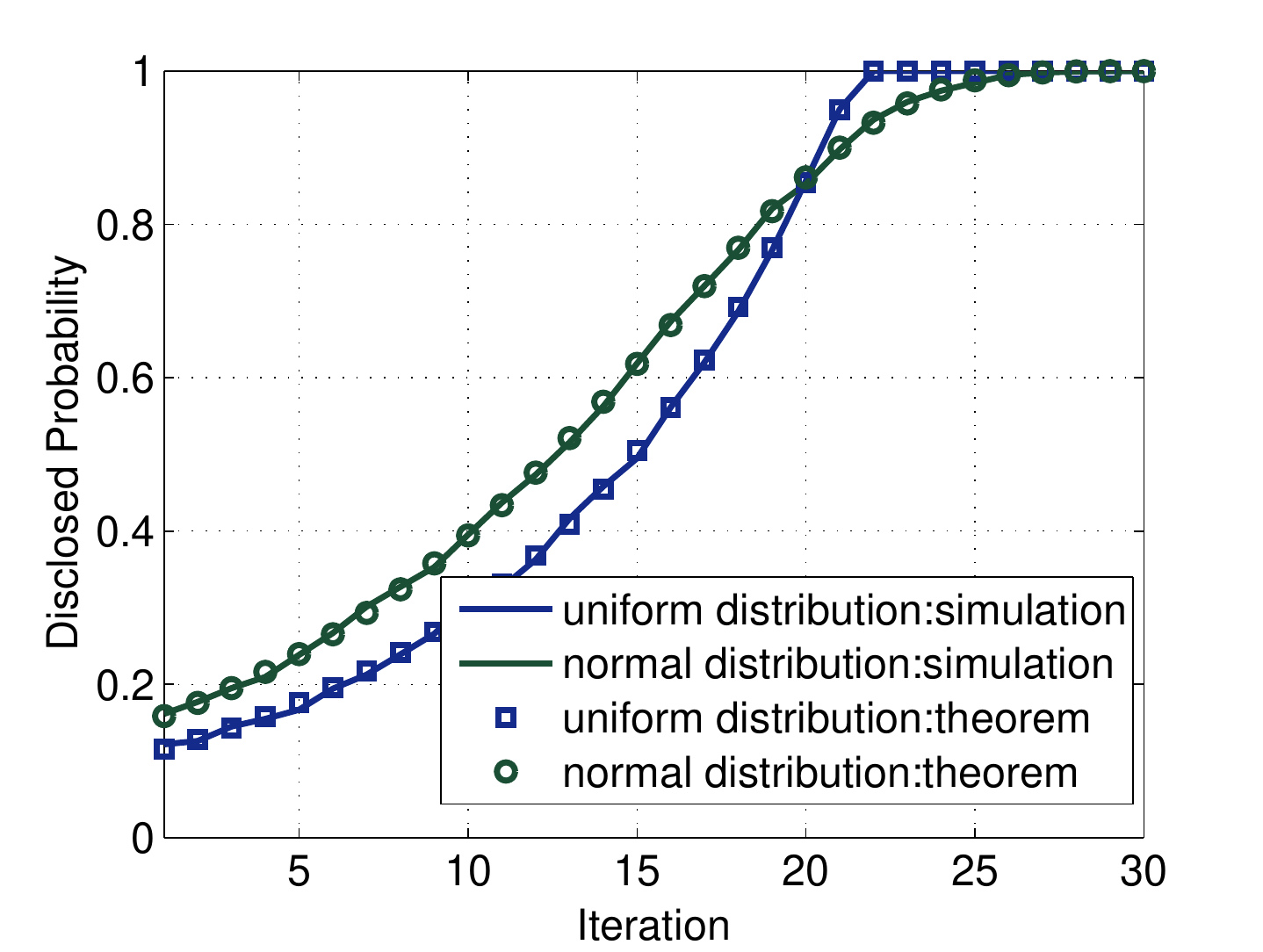}}
\caption{The convergence and privacy comparison under different random noise distribution.}\label{pricepay}
\end{center}
\vspace*{-0pt}
\end{figure*}

Fig. \ref{pricepay}(a) compares the convergence speed of the  PPAC algorithm using normal and uniform distribution noises, in which the basic design is the same as PPAC proposed in \cite{yilin15tac}. It is observed that under the two different distributions, the PPAC algorithm has the same convergence speed. This justifies that the convergence speed only depends on the eigenvalues of the weighted matrix $W$ and the value of $\varrho$ as proved in \cite{yilin15tac}.

Fig. \ref{pricepay}(b) compares the $(\epsilon, \delta)$-data-privacy under $\mathcal{I}_i^0$ with normal and uniform distribution noises. In simulation, we conduct $10,000$ simulation runs. For each run, one node first generates a state $\theta_i(0)$ randomly with the given distribution, and the other node generates $10,000$ random numbers with the same distribution and use them as the estimation of $\theta_i(0)$ (i.e., $\hat{\theta}_i(0)$). Then, one get the probability of $|\hat{\theta}_i(0)-{\theta}_i(0)|\leq\epsilon$ in each run, and we use the maximum probability among these in all runs simulation as the value of $\delta$. For the theoretical results, we use (\ref{deltainq})  to calculate the value of $\delta$ under two different distributions. Clearly, one can observe from Fig. \ref{pricepay}(b) that uniform distribution is much better than normal distribution in the sense of $(\epsilon, \delta)$-data-privacy. It is also observed that $\delta$ in simulation matches its value in theory. 

Fig. \ref{pricepay}(c) compares the $(\epsilon, \delta)$-data-privacy under $\mathcal{I}_i^1$  using normal and uniform distribution noises. The simulations here are conducted similarly as those in Fig. \ref{pricepay}(b), except that when the iteration increases, the variance of the noises will be changed to $\sigma^2=\varrho^{2k}$ since (\ref{pri1k}) will be used for estimation at iteration $k$. We use (\ref{deltainq})   to calculate the value of $\delta$, and the corresponding results are denoted as theoretical results.  Both in simulation and theory, we set $\epsilon=0.2$. As shown in Fig. \ref{pricepay}(c), the maximum disclosure probability increases  with iteration and will converge to $1$, i.e., the privacy decays with iteration and will eventually be compromised. 

\subsection{Evaluation}
In this subsection, we will evaluate the performance of the OPAC algorithm. Using the same setting as the above subsection, the OPAC algorithm can guarantee the similar privacy as the blue line shown in Fig. \ref{pricepay}(b) under $\mathcal{I}_i^1$. This is because uniform distribution noise is used in OPAC and the secret function makes the subsequent ($k\geq 1$) information cannot increase the disclosure probability. Therefore, the OPAC can guarantee much stronger privacy than the GPAC, since it can achieve the same data-privacy under $\mathcal{I}_i^1$ as the GPAC under $\mathcal{I}_i^0$.
\begin{figure}[http]
\begin{center}
\vspace*{-5pt}
\subfigure[converge to average consensus]{\label{opaclowpr}
\includegraphics[width=0.4\textwidth]{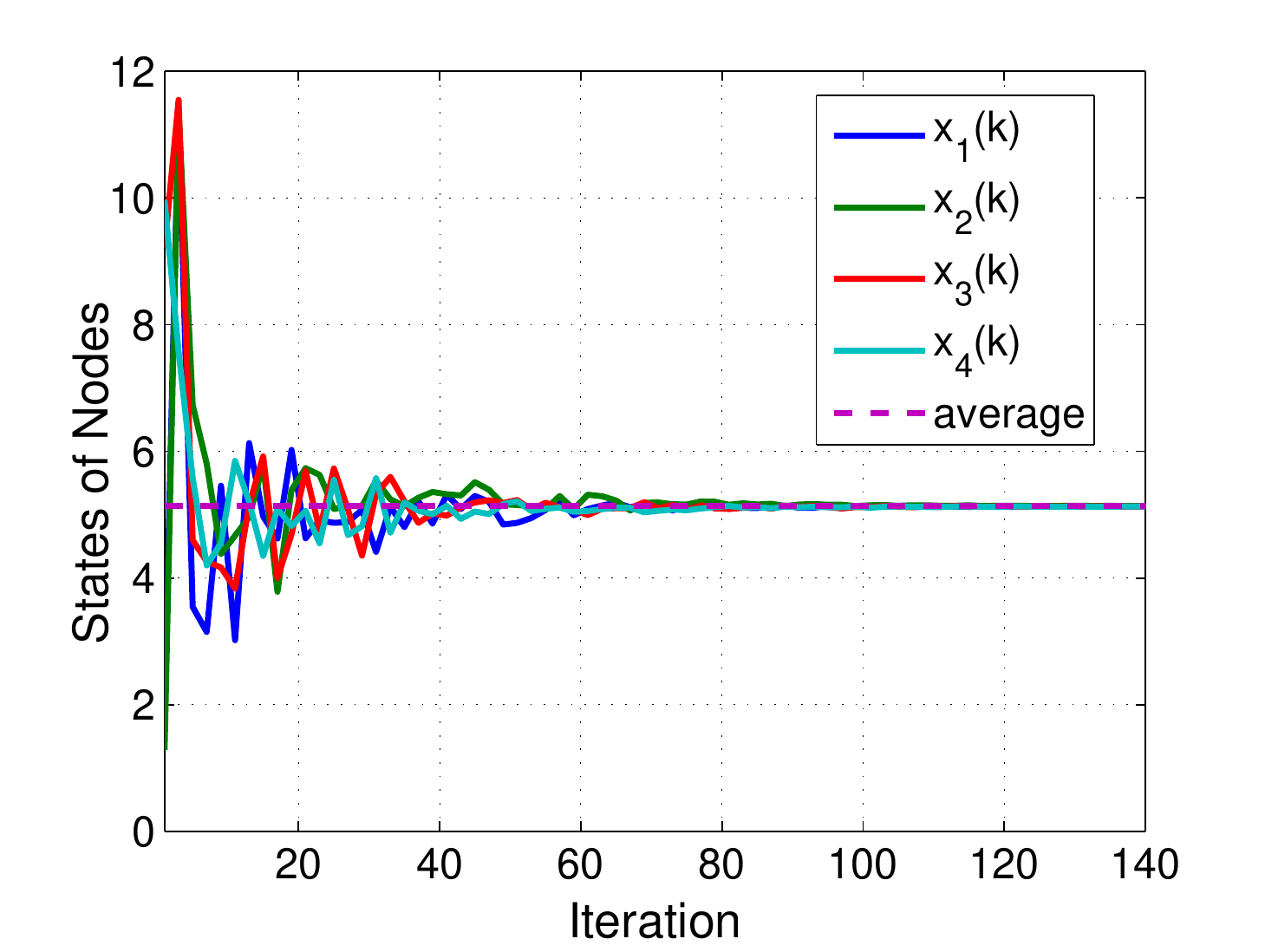}} \subfigure[same covergence speed]{\label{opacpayment}
\includegraphics[width=0.4\textwidth]{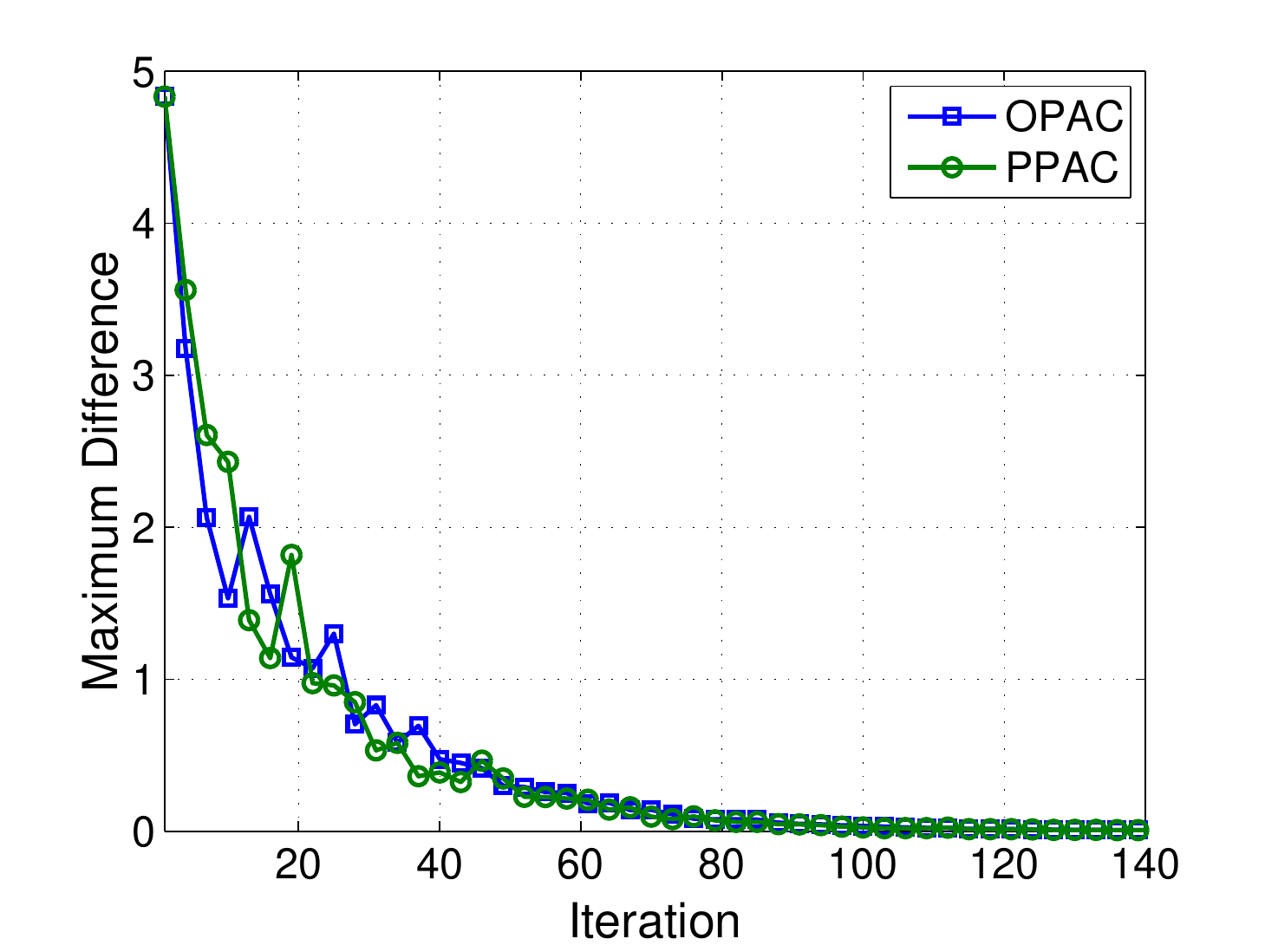}}
\caption{The performance evaluation of the OPAC algorithm.}\label{opacpricepay}
\end{center}
\vspace*{-5pt}
\end{figure}

Then, we test the convergence of the OPAC algorithm. Set $F_{ij}={i+2j\over 50}$. As shown in Fig. \ref{opaclowpr}, we find that the nodes' states will converge to the exact average with the OPAC, which means that an exact average consensus can be achieved by the proposed algorithm. Fig. \ref{opacpayment} compares the convergence speed of the OPAC and PPAC, it is found that they almost have the same convergence speed. Hence,  added secret function will not affect the convergence speed.

\section{Conclusions}\label{sec:conclusions}

In this paper, we investigated the privacy of the GPAC algorithm. We proposed a novel privacy definition, named $(\epsilon, \delta)$-data-privacy, to depict the relationship between privacy and estimation accuracy, so that the degree of the privacy can be well quantified. We proved that the GPAC algorithm achieves $(\epsilon, \delta)$-data-privacy, and obtained the  closed-form expression of the relationship between $\epsilon$ and $\delta$. We also proved that the noise with uniform distribution guarantee a highest privacy when $\epsilon$ is small enough.  We revealed that the privacy will be lost when the information used in each consensus iteration is available to the other nodes. Then, to solve this problem and achieve highest $(\epsilon, \delta)$-data-privacy, we proposed OPAC algorithm, followed by the convergence and privacy analysis.  Lastly, simulations are conducted to demonstrate the efficiency of the proposed algorithm.

\end{document}